\documentclass[Review,sagev,times]{sagej}

\pdfoutput=1 

\usepackage{moreverb,url}

\usepackage[colorlinks,bookmarksopen,bookmarksnumbered,citecolor=red,urlcolor=red]{hyperref}

\newcommand\BibTeB{{\rmfamily B\kern-.05em \textsc{i\kern-.025em b}\kern-.08emT\kern-.1667em\lower.7ex\hbox{E}\kern-.125emB}}

\DeclareMathOperator{\sgn}{sgn}
\DeclareMathOperator{\se}{se}
\newtheorem{theorem}{Theorem}
\newtheorem{proposition}{Proposition}

\def\volumeyear{2018}

\begin{document}

\runninghead{van Zwet}

\title{A default prior for regression coefficients}

\author{Erik van Zwet\affilnum{1}}

\affiliation{\affilnum{1}Leiden University Medical Center}

\corrauth{Erik van Zwet,
Department of Biomedical Data Sciences, 
Leiden University Medical Center,
The Netherlands.}

\email{E.W.van\_Zwet@lumc.nl}

\begin{abstract}
When the sample size is not too small, M-estimators of regression coefficients are approximately normal and unbiased. This leads to the familiar frequentist inference in terms of normality-based confidence intervals and $p$-values. From a Bayesian perspective, use of the (improper) uniform prior yields matching results in the sense that posterior quantiles agree with one-sided confidence bounds. For this, and various other reasons, the uniform prior is often considered objective or non-informative. In spite of this, we argue that the uniform prior is not suitable as a default prior for inference about a regression coefficient in the context of the bio-medical and social sciences. We propose that a more suitable default choice is  the normal distribution with mean zero and standard deviation equal to the standard error of the M-estimator. We base this recommendation on two arguments. First, we show that this prior is non-informative for inference about the sign of the regression coefficient. Secondly, we show that this prior agrees well with a meta-analysis of 50 articles from the MEDLINE database.
\end{abstract}

\keywords{objective prior, Jeffreys prior, objective Bayes, empirical Bayes, type S error, type M error, p-value debate, normal-normal model}

\maketitle

\section{Introduction}
Regression modeling plays a central role in the bio-medical and social sciences. Linear and generalized linear models (with and without random effects), generalized estimating equations (GEE) and quantile regression offer great flexibility and are easy to use. Also, simple two group comparisons can usually be viewed in terms of a regression model. When the sample size is not too small, the statistical analysis can be based on the fact that M-estimators of regression coefficients are approximately normal and unbiased \cite{stefanski2002calculus}. This leads to the familiar (frequentist) inference in terms of normality-based confidence intervals and $p$-values.

We will avoid small sample issues by assuming that we have a normally distributed, unbiased estimator $B$ of a regression coefficient $\beta$ with known standard error $\se$.  Then we have for any fixed $\beta$
\begin{equation}\label{conf}
P_\beta(B- 1.96 \se < \beta < B + 1.96 \se)=0.95.
\end{equation}
While our set-up may appear to be overly simplistic, we emphasize that inference about regression parameters based on Wald type confidence intervals (and associated $p$-values) is very common throughout the life sciences. Exact intervals based on the $t$-distribution are available in linear models, but the difference is already very small when the degrees of freedom exceed, say, 40.

Statement  (\ref{conf}) describes the long-run coverage performance of the random interval $[B- 1.96 \se ,B + 1.96 \se]$. However, this confidence statement is often mistakenly interpreted as the conditional probability statement
\begin{equation}\label{prob}
P(B - 1.96 \se < \beta < B + 1.96 \se \mid B=b)=0.95,
\end{equation}
where $b$ is the observed value of the estimator and $\beta$ is viewed as a random variable. We refer to Greenland et al.\ \cite{greenland2016statistical} for a discussion of this misinterpretation. Statement (\ref{prob}) is arguably more relevant than (\ref{conf}) because it refers to the data at hand, rather than the procedure being used. This may explain, at least in part, the pervasiveness of the misinterpretation; it is what researchers want to know.

Statement (\ref{prob}) is actually only valid if we assume that $\beta$ has the (improper) uniform or ``flat" prior distribution. This has lead some authors to consider the uniform prior to be an objective or non-informative prior \cite{ghosh2011objective}. Many other criteria have been proposed by which a prior may be considered to be objective \cite{kass1996selection}, but in the normal location model with known standard deviation they all yield the (improper) uniform distribution as the unique objective prior. 

We find that the flat prior is used  for Bayesian inference about regression coefficients in two distinct situations. It is used explicitly with the goal of objective Bayesian inference\cite{berger2006case} and implicitly whenever the confidence interval for a regression coefficient is interpreted as a credibility interval. 

This paper consists of two parts. In the first part (section \ref{objective}) we discuss the objective Bayesian approach. Loosely speaking, the goal of this approach is to minimize the influence of the prior on the posterior. However, this depends on which aspect of the posterior we are considering. While the flat prior may well be considered non-informative for $\beta$, it is very informative {\em both} for the magnitude and the sign of $\beta$.  This is just a consequence of the fact that a very diffuse prior favors large values of $|\beta|$. Hence, use of the flat prior implies that the magnitude of $\beta$ will be inflated and the evidence about its sign will be exaggerated

This is problematic, because for regression coefficients (other than the intercept) we are typically most interested in the sign and the magnitude. In section \ref{par interest} we go one step further and argue that the sign is often of primary interest. We start from the premise that associations studied in the life sciences are almost never exactly zero and that it is therefore of special importance to quantify the evidence for the {\em direction} of a certain association. From a Bayesian perspective, this means that we are primarily interested in $P(\beta >0 \mid B)$. 

Now, if we want to avoid undue influence of the prior on $P(\beta >0 \mid B)$, it is natural to use a prior for $\beta$ such that $P(\beta >0 \mid B)$  has the standard uniform distribution. Theorem 1 asserts that the normal distribution with mean zero and standard deviation equal to the standard error of the unbiased estimator is such a prior. In other words, this prior may be considered to be non-informative for inference about the sign of $\beta$.

In the second part of the paper (section \ref{empirical}) we turn to the Empirical Bayesian approach\cite{carlin2010bayes} \cite{efron2012large}. MEDLINE is an extensive bibliographic database of life sciences and biomedical information. Compiled by the United States National Library of Medicine, it is freely available on the Internet and searchable via PubMed. In the absence of additional prior information, we may consider papers from MEDLINE to be {\em exchangeable}. This implies that we can use a sample of MEDLINE papers to estimate a suitable prior distribution for regression coefficients. Such an estimated prior is objective in the sense that it is essentially free of any personal opinions or biases. Based on a sample of 50 MEDLINE articles, we estimate the distribution of regression coefficients (other than the intercept) to be normal with mean zero and standard deviation $1.28\,\se$.

Objective Bayes and Empirical Bayes both aim for objectivity, but from very different points of view. It is therefore quite remarkable that both approaches lead to such similar priors. This supports the main conclusion of this paper that the flat prior is not suitable for objective inference about regression coefficients. 

In section \ref{results} we discuss several reasons why the factor of 1.28 is likely to be an overestimate. Therefore, we recommend that the normal distribution with mean zero and standard deviation $\se$ is a suitably conservative default prior. This prior combines the theoretical support of section  \ref{par interest} with the empirical evidence from section \ref{pubmed}.  We do stress that this default prior is not always appropriate. In section \ref{discussion} discuss several situations in which it is {\em not} to be used.

\begin{figure}[htp] \centering{
\includegraphics[scale=0.7]{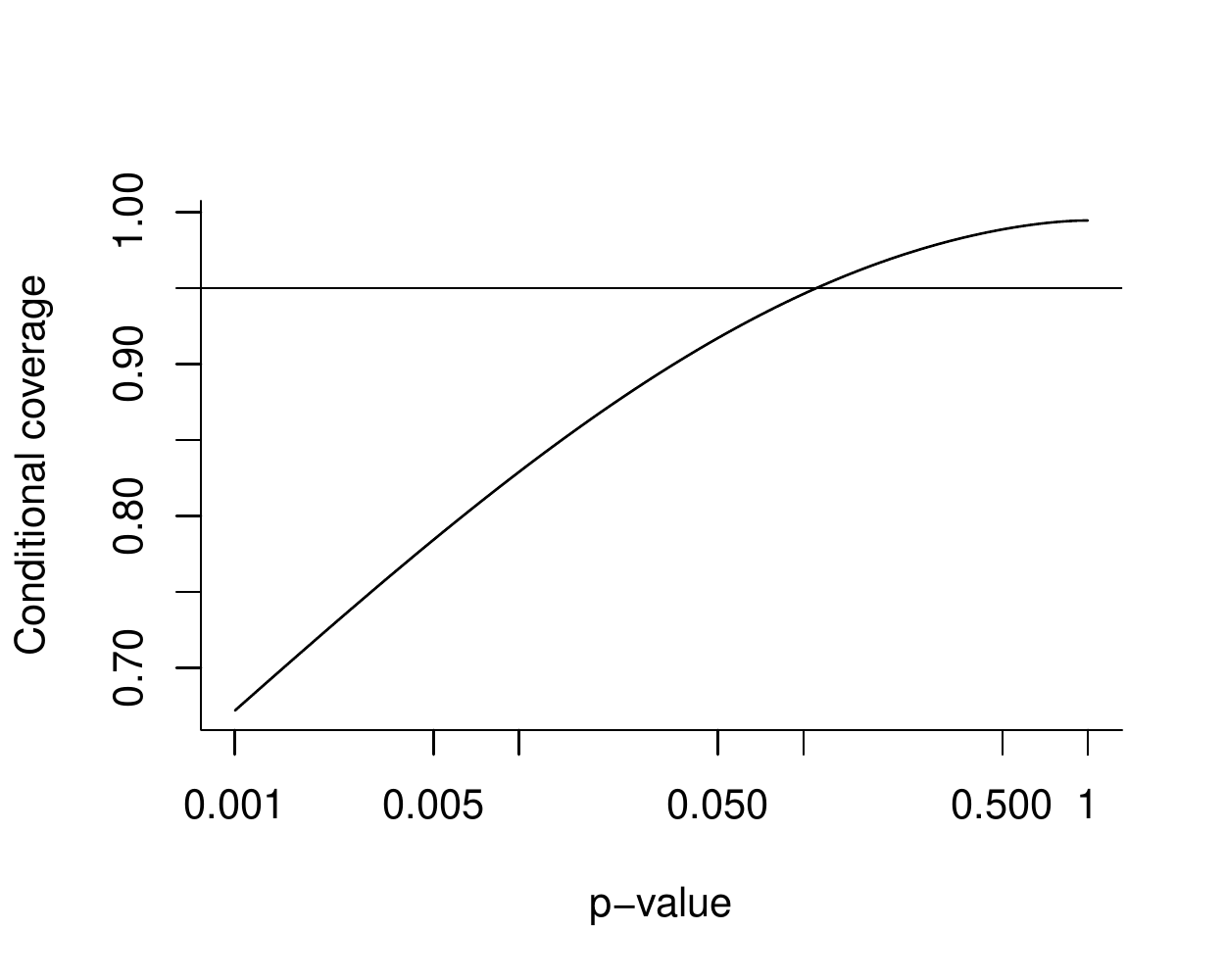}
\caption{Conditional coverage when $\beta$ has the normal prior with mean zero and standard deviation $\se$, as a function of the two-sided $p$-value.}\label{fig:coverage}
}
\end{figure}

Upon observing $B=b$, the resulting posterior is the normal distribution with mean $b/2$ and standard deviation $\se/\sqrt{2}$.  Since statement (\ref{conf}) holds for all $\beta$,  it remains valid under the proposed prior. However,  the conditional coverage of the usual confidence interval becomes
\begin{align}\label{prob2}
P(B - 1.96 \se &< \beta < B + 1.96 \se \mid B=b) \notag \\
&=\Phi\left( \frac{b}{\sqrt{2} \se} + 1.96 \sqrt{2} \right) - \Phi\left( \frac{b}{\sqrt{2} \se} - 1.96 \sqrt{2} \right),
\end{align}
where $\Phi$ is the standard normal cumulative distribution function.  We show the conditional coverage in Figure \ref{fig:coverage}. Instead of $b$, we put the two-sided $p$-value  $2\Phi(-|b|/\se)$ on the $x$-axis to emphasize that with increasing significance, the coverage probability decreases. Also, we have
\begin{equation}\label{prob3}
P \left(\frac{B}{2} - 1.96 \frac{\se}{\sqrt{2}} < \beta < \frac{B}{2} + 1.96 \frac{\se}{\sqrt{2}} \mid B=b \right)=0.95,
\end{equation}
and
\begin{equation}
P(\beta > 0 \mid B=b)= \Phi(b/\sqrt{2}\se).
\end{equation}
Finally, $P(\beta < 0 \mid B=b)=1-P(\beta > 0 \mid B=b)$.

We note that several other authors\cite{berger1999default} \cite{gelman2000type}\cite{senn2007trying} \cite{gelman2008weakly} \cite{seaman2012hidden} have warned against using the flat prior as a default choice and some of those authors have proposed priors that are qualitatively similar to ours, in the sense of being symmetric and concentrated around zero.

\section{Objective Bayes}\label{objective}
The Bayesian method offers a principled and coherent approach to statistical inference, but it does require the specification of a prior distribution. If one is unable or unwilling to formalize external information into a prior, it is tempting to use a so-called objective (or non-informative) prior to try to avoid undue influence on the posterior \cite{berger2006case}.

Many criteria have been proposed by which a prior may be considered to be objective \cite{kass1996selection}. In the normal location model with known standard deviation they all point to the flat prior. Here we review two important criteria; one based on invariance of the statistical decision problem and one based on the principle of parameterization invariance. We will argue that neither criterium is as compelling as it may seem. An information theoretic argument in section \ref{par interest} leads us to propose the normal distribution with mean zero and standard deviation equal to the standard error of the unbiased estimator as a suitable objective prior.

\subsection{Invariant decision problems}\label{invariant}
The statistical decision problem of estimating the mean $\beta$ of a normal distribution under squared error loss is invariant under translation \cite{berger2013statistical}. In the absence of prior information, it is sensible that the prior distribution on $\beta$ should be translation-invariant because that will imply that the Bayes rule is translation-invariant as well \cite{berger2013statistical}. The uniform distribution is the only such distribution over the real numbers (up to a multiplicative constant).

Applied researchers are often particularly interested in the sign or the magnitude of a regression coefficient. This focus may be reflected formally by a loss function that depends on $\beta$ only through its sign or magnitude. Under such loss functions, the decision problem is no longer invariant and the above argument no longer applies.

Suppose we want to estimate the magnitude of $\beta$. Since $B$ is unbiased for $\beta$, it follows by Jensen's inequality that $|B|$ is positively biased for $|\beta|$. For fixed $\beta$, $|B|$ has the folded normal distribution with mean
\begin{equation}
E_\beta |B| = |\beta| + \sqrt{\frac{2}{\pi}} \se e^{-\beta^2/2\se^2}  - 2 |\beta| \Phi\left( - \frac{|\beta|} {\se} \right).
\end{equation}
The bias $E_\beta|B| - |\beta|$ is maximal at $\beta=0$ when it is equal to $\sqrt{2/\pi}\, \se \approx 0.8\, \se$.  Now, if we use the flat prior, then the posterior distribution of $\beta$ given $B=b$ is normal with mean $x$ and standard deviation $\se$. Hence, the posterior distribution of $|\beta|$ is the folded normal with mean
\begin{equation}
E(|\beta| \mid B=b) = |b| + \sqrt{\frac{2}{\pi}} \se e^{-b^2/2\se^2}  - 2 |b| \Phi\left( - \frac{|b|} {\se} \right)
\end{equation}
This posterior mean is the Bayes estimator of $|\beta|$ under squared error loss. We see that it is even larger than $|B|$, which is already positively biased for $|\beta|$. Evidently, the flat prior is very informative for $|\beta|$.

Similar problems arise with inference about the sign of $\beta$.  If we use the flat prior then
\begin{equation}
P(\sgn(\beta) = \sgn(B) \mid B=b) = \Phi\left( \frac{|b|}{\se} \right).
\end{equation}

\noindent
Now, the unimodal, symmetric priors are a natural class to consider for objective inference about the sign of $\beta$. The following proposition is essentially due to Casella and Berger\cite{casella1987reconciling}. 

\begin{proposition}
Suppose $\beta$ has a prior distribution $\pi$ which has a unimodal density and is symmetric about zero. Also, suppose that conditionally on $\beta$, $B$ has the normal distribution with mean $\beta$ and standard deviation $\se>0$. For any $b$, 
\begin{equation}
P_\pi(\sgn(\beta) = \sgn(B) \mid B=b) \leq \Phi\left( \frac{|b|}{\se} \right).
\end{equation}
\end{proposition}

The proposition asserts that the {\em perceived} evidence that $\beta$ has the same sign as $B$ is maximal under the uniform prior among all unimodal, symmetric priors. We conclude that the flat prior is very informative for inference about the sign. This issue is has long been recognized and was discussed, for instance, by Berger and Mortera \cite{berger1999default}.

\subsection{Parameterization invariance}\label{jeffreys}
Parameterization invariance refers to the principle that the construction of a non-informative prior should not depend on the parameterization that happens to have been chosen. This principle leads to Jeffreys rule \cite{jeffreys1946invariant} which is to define the prior as (proportional to) the square root of the Fisher information. This construction is unaffected by smooth reparameterization.  In the normal location problem with known standard deviation, Jeffreys prior is the uniform distribution. We will now present a non-smooth reparameterization that does affect Jeffreys rule. 

\begin{figure}[htp] \centering{
\includegraphics[scale=0.8]{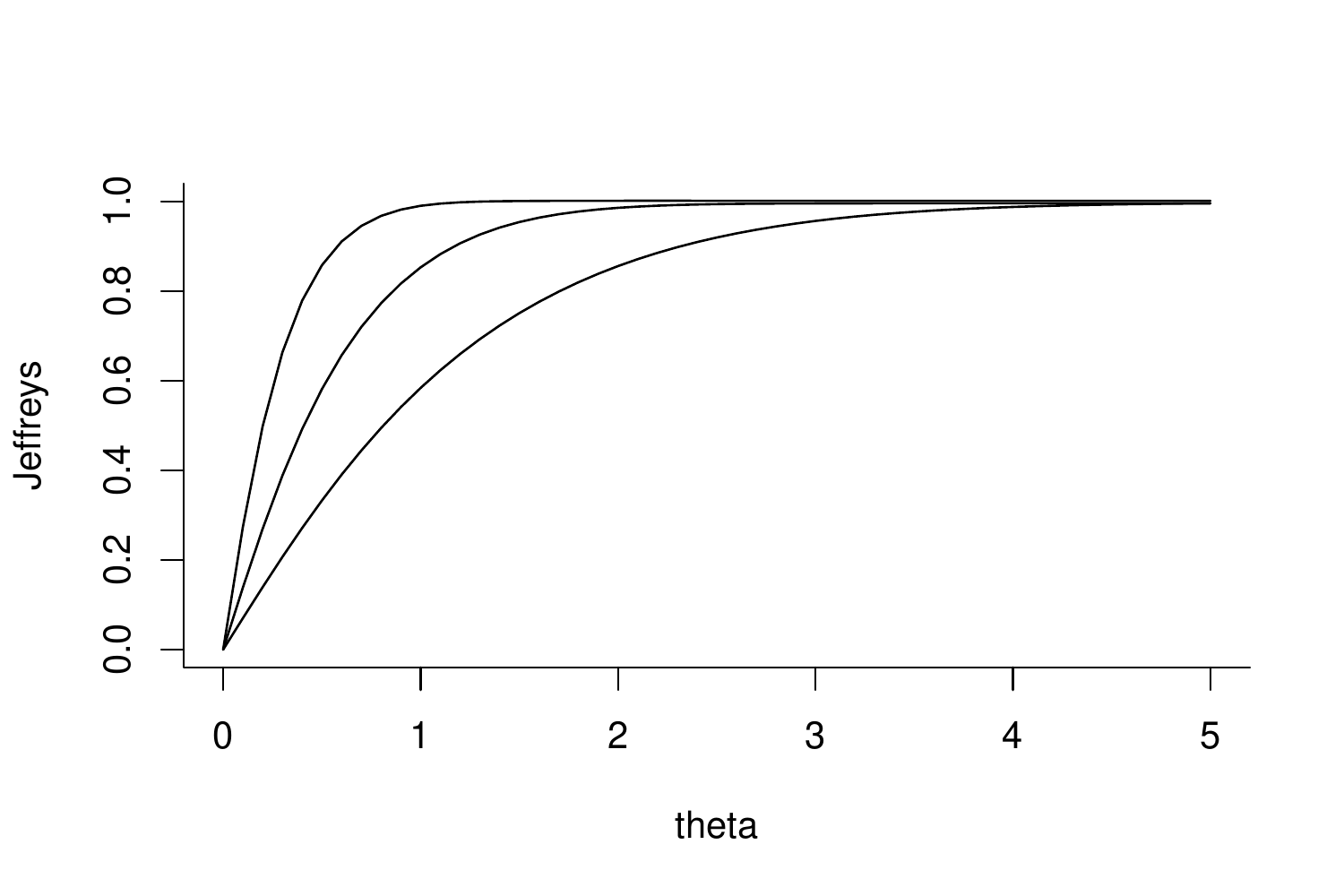}}
\caption{Jeffreys prior for $\theta=|\beta|$ when $\se=0.5, 1, 2$ (from left to right).}\label{fig:jeffreys}
\end{figure}

We can reparameterize the location  problem quite naturally in terms of the sign and  absolute value of $\beta$. This reparameterization is of course not smooth and the Fisher information is not defined because the sign is discrete. However, it is natural to put the Bernoulli(1/2) prior on the sign. Writing $\theta=|\beta|$, the distribution of $B$ becomes a two-component mixture with density
\begin{equation}\label{lik}
f(b \mid  \theta) = \frac{1}{2 \se} \varphi \left(\frac{b + \theta}{\se} \right) + \frac{1}{2 \se} \varphi \left(\frac{b - \theta}{\se} \right),
\end{equation}
where $\varphi$ is the standard normal density function. Jeffreys prior on $\theta$  is defined as
\begin{equation} \label{jeffreys prior}
f(\theta) \propto \left( E_\theta \left( \frac{\partial}{\partial \theta} \log f(b \mid  \theta) \right)^2 \right)^{1/2},
\end{equation} 
where
\begin{align}
& \frac{\partial}{\partial \theta} \log f(b \mid  \theta) \\ \notag
& \propto \frac{-(b+\theta)\varphi((b + \theta)/\se) + (b-\theta) \varphi((b - \theta)/\se)}{\varphi((b + \theta)/\se) + \varphi((b - \theta)/\se)}.
\end{align} 
A closed form expression for (\ref{jeffreys prior}) is not available, but we can easily compute it. We show Jeffreys prior for various values of $\se$  in Figure \ref{fig:jeffreys}. We see that this prior is no longer uniform. In fact, it is even more widely dispersed.

\subsection{The parameter of interest}\label{par interest}
Almost all criteria that have been put forward for constructing a non-informative prior agree on the uniform distribution for a location parameter. However, this prior leads to overestimation of the magnitude and overconfidence about the sign. Here, we take a different approach.

Much research in the bio-medical and social sciences is aimed at assessing the evidence about the association between two variables, often after correcting for additional variables. The inferential  approach that is most often used, is to perform a hypothesis test. Hypothesis testing is ubiquitous, but it is not without criticism. We refer to the American Statistical Association's statement \cite{wasserstein2016asa} for a discussion of the issues involved in the so-called ``$p$-value debate''. Much of the criticism of hypothesis testing centers on the fact that the $p$-value is often misinterpreted as the posterior probability that the null hypothesis of no association is true. 

In many situations, it is unlikely that $\beta$ is {\em exactly} zero. If one assumes from the outset that $\beta$ is probably not exactly zero, then attention naturally shifts to inference about the sign of $\beta$ \cite{gelman2000type}. John Tukey has argued this point particularly forcefully: ``All we know about the world teaches us that the effects of A and B are always different -- in some decimal place -- for any A and B. Thus asking ``Are the effects different?'' is foolish. What we should be answering first is ``Can we tell the direction in which the effects of A differ from the effects of B?'' In other words, can we be confident about the direction from A to B?''\cite{tukey1991philosophy}

Following this reasoning, we propose that the (data-dependent) parameter $P(\beta > 0 \mid B)$ is of primary importance in much empirical research.  If indeed  $P(\beta > 0 \mid B)$ is of primary interest and if one wants to avoid undue influence of the prior, then it is quite natural to put the uniform prior on this parameter \cite{bayes1763essay} \cite{laplace2012pierre} \cite{jaynes1968prior}.  It is not immediately obvious which prior on $\beta$ implies the uniform prior on $P(\beta > 0 \mid B)$, but we have the following connection between the uniform and the normal distribution, which is new as far as we know.
\begin{theorem}
Suppose  the prior distribution of $\beta$ is normal with mean $0$ and standard deviation $\se$ and conditionally on $\beta$, $B$ has the normal distribution with mean $\beta$ and standard deviation $\se$. Then $P(\beta > 0 \mid B)$ has the standard uniform distribution.
\end{theorem}

We believe that $N(0,\se^2)$ is actually the only prior for $\beta$ such that $P(\beta > 0 \mid B)$ has the uniform distribution, but we have not been able to prove it. It is of course unique among all normal distributions.  Now, on the basis of Theorem 1, we propose that the $N(0,\se^2)$ prior on $\beta$ is appropriate for objective Bayesian inference about the sign of $\beta$. 

It is interesting to mention that  our proposal would follow from an application of the reference method of Bernardo \cite{berger2009formal} to the parameter $P(\beta > 0 \mid B)$. However, it should be stressed that the formal definition of the reference prior does not cover data dependent parameters.

Recall that $\se$ represents the standard error of the estimator $B$ and therefore depends on the sample size. Consequently, the proposed prior depends on the sample size as well. This may seem awkward from a subjective Bayesian point of view and requires justification. Now, objective Bayesian inference aims to avoid undue influence of the prior on the posterior. Since the prior exerts its influence on the posterior through the likelihood, the choice of objective prior will often depend on the likelihood. While this may be at odds with the subjective Bayesian point of view, it is inherent in the goal of objectivity  \cite{gelman2017prior}. 

A researcher's prior beliefs about $\beta$ are surely not informed by the sample size, but it does work the other way around. It is typically the case that the researcher's prior beliefs influence the sample size through (formal or informal) sample size calculations. Therefore, it is not unreasonable for the reader of a bio-medical paper to take the sample size into account in his or her prior beliefs about $\beta$.

\section{Empirical Bayes}\label{empirical}
The fact that we are considering a regression coefficient  in the context of the bio-medical or social sciences may in itself be taken as prior information. Here, we will use external information from the the extensive MEDLINE database to inform our prior beliefs, but do so in an objective manner. In the absence of additional prior information, we consider papers from MEDLINE to be {\em exchangeable}. This implies that we can use a sample of MEDLINE papers to estimate a suitable prior. Estimating the prior is often referred to as Empirical Bayes \cite{carlin2010bayes} \cite{efron2012large}.

Let $\beta_{ij}$ denote the $i$-th  regression coefficient in paper $j$ and assume the following hierarchical model.
\begin{itemize}
\item $\phi_j$ has the normal distribution with mean $\phi$ and variance $\sigma^2$
\item Conditionally on $\phi_j$, $\beta_{ij}$ has the normal distribution with mean zero and variance $\phi_j\se_{ij}^2$
\item Conditionally on $\beta_{ij}$, $B_{ij}$ has the normal distribution with mean $\beta_{ij}$ and variance $\se_{ij}^2$
\end{itemize}
It follows that conditionally on $\phi_j$, the ``$z$-value'' $Z_{ij}=B_{ij}/\se_{ij}$ has the normal distribution with mean zero and variance $\phi_j +1$. The idea behind the model is that some studies have a larger sample size or a more predictable outcome than others, and that will lead to $z$-values of larger magnitude. To capture this in our model, we have the variance of the $z$-values depend on the study. 

Now, conditionally on $\phi_j$, $Z_{ij}^2$ has the Gamma distribution with mean $\phi_j +1$ and shape $1/2$. Therefore, based on a sample of $z$-values, we can estimate the parameters of this model by fitting a generalized linear mixed model with Gamma distribution, shape $1/2$, identity link and Gaussian random effect per study, to the squared $z$-values. If we use an offset of one, then the intercept of this model estimates $\phi$. Note that the prior we proposed in section \ref{par interest} for objective inference about the sign has $\phi=1$.

We will now describe how we collected our data. 

\subsection{MEDLINE}\label{pubmed}
It is well-known that for various reasons (p-hacking, fishing, file drawer effect, etc.) reported effects tend to be inflated \cite{rothstein2006publication} \cite{ioannidis2005contradicted} \cite{ioannidis2008most} \cite{button2013power} \cite{gelman2013garden}. We have tried to collect ``honest'' effects as follows. It is a fairly common practice in the life sciences to build multivariate regression models in two steps. First, the researchers run a number of univariate regressions for all predictors that they believe could have an important effect. Next, those predictors with a $p$-value below some threshold are selected for the multivariate model. While this approach is statistically unsound, we believe that the univariate regressions should be largely unaffected by selection on significance, simply because that selection is still to be done!

We entered the search term ``univariate multivariate regression'' into the PubMed system which yielded over 20,000 results. We selected the 80 most recent (in August 2018), consecutive results. Of these, 50 reported univariate and multivariate $p$-values. The 30 other papers were either unavailable under the license of the University of Leiden, or did not report $p$-values precisely (summaries such as $p<0.01$ or $p<0.05$ or ``NS" are common).  In case results were reported for more than one outcome, we used only the first. We collected in total 576 univariate $p$-values (all two-sided). We did not collect $p$-values for the intercept, or $p$-values based on $F$ tests. Also, we did not collect $p$-values below 0.001. There are two reasons for that. First, because $p$-values below 0.001 are usually only reported as such. Second, because we would consider such a small $p$-value as evidence against our default prior and would not recommend its use in those cases, c.f.\ section \ref{discussion}. All data are available as a supplement to this article.

\begin{figure}[htp] \centering{
\includegraphics[scale=0.8]{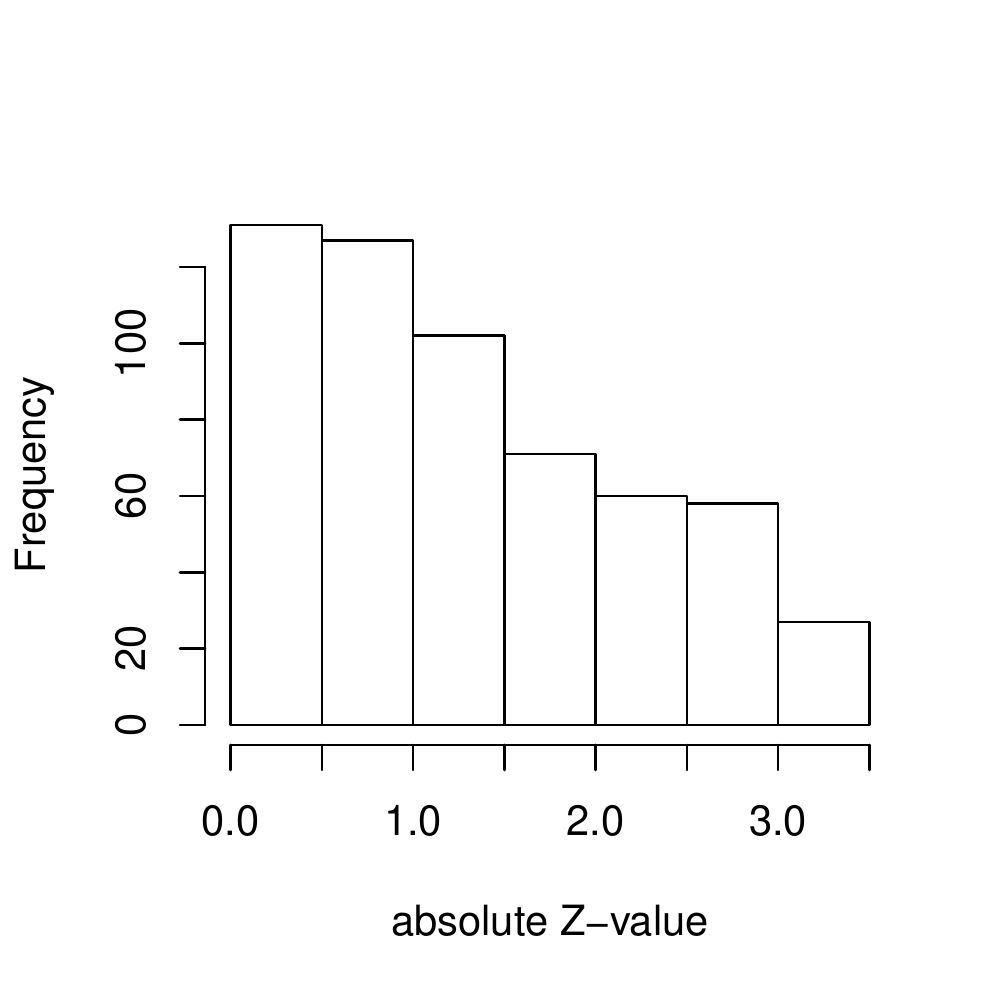}}
\caption{Histogram of 576 absolute $z$-values from 50 MEDLINE articles. Restricted to $[0,3.29]$.}\label{fig:zvalues}
\end{figure}

We converted the two-sided $p$-values to absolute $z$-values through $|z| = |\Phi^{-1}(p/2)|$ and display the result in Figure \ref{fig:zvalues}. The fact that we do not have the sign of the $z$-values is not relevant for our purpose, because we consider only symmetric distributions as plausible default priors. 

\subsection{Results}\label{results}
We have fitted the Gamma regression model as described above to 576 squared $z$ -values obtained from 50 MEDLINE articles. We are primarily interested in $\sqrt{\phi}$ which is the relation between the standard deviation of the prior and the standard error of the unbiased estimator. We estimate $\sqrt{\phi}$ to be 1.28 (95\% CI: 1.12 to 1.42). The details of our analysis including the {\tt R} code are in the supplemental material. In addition to the (conditional) mixed model, we also fitted a marginal model. As the results were very similar, we do not report them here. They are included in the supplemental material.

Our proposal in section \ref{par interest} for an objective prior for inference about the sign has  $\sqrt{\phi}=1$, while we estimate $\sqrt{\phi}=1.28$ for a typical study. We consider this to be quite close, especially compared to the flat prior. Moreover, the value 1.28 is likely to be an over-estimate, because it is safe to assume that even the univariate $z$-values we have collected are inflated. For  example, researchers may have dichotomized certain predictor variables in a favorable way, or decided not to report predictors that showed no association with the outcome at all.

\section{Discussion}\label{discussion}
The flat prior is used  for Bayesian inference about regression coefficients in two distinct situations. It is used explicitly with the goal of objective Bayesian inference\cite{berger2006case} and implicitly whenever the confidence interval for a regression coefficient is interpreted as a credibility interval. This is problematic, because the uniform distribution is not realistic at all in the context of bio-medical research. Consequently, its use leads to overestimation of the magnitude of the regression coefficient and overconfidence about its sign. 

In this paper, we have proposed a different prior to be used as a default. Suppose we have an unbiased, normally distributed estimator $B$ of $\beta$ with standard error $\se$. Then we have argued that the normal distribution with mean zero and standard deviation $\se$ is more suitable as a default prior than the uniform distribution. Equivalently, upon observing $B=b$, the normal distribution with mean $b/2$ and standard deviation $\se/\sqrt{2}$ is a more suitable default posterior than the normal distribution with mean $b$ and standard deviation $\se$.

We based our proposal on an information theoretic argument (Theorem 1), but also demonstrated that our prior agrees quite well with data about regression coefficients we gathered from 50 papers from the bio-medical and social sciences. We do want to stress that our default prior is not meant to be a universal prior.  There are, at least, four circumstances when our proposed default prior should {\em not} be used. 
\begin{enumerate}
\item Our prior is concentrated around zero which is usually not appropriate for the intercept as there is no reason a priori why the intercept should be close to zero. 
\item A different prior should be used when additional external prior information is available. We do feel, however, that it is the responsibility of the researcher to convince the reader (i.e.\ the scientific community) that his or her study is different from a typical MEDLINE study.
\item Our proposed prior should not be used if a two-sided $p$-value less than 0.001 is observed. This is a form of prior-data conflict \cite{evans2006checking}, because such a small $p$-value is quite unlikely under our prior. To be precise, a two-sided $p$-value less than 0.001 corresponds to a $z$-value exceeding 3.29 in absolute value, and the probability of that event is about 2\% under our prior. 
\item Our proposed prior should not be used in situations in high dimensional situations where we can use empirical Bayes methods to reliably estimate a prior that is specific to the study under consideration. Many examples of such situations are discussed in a book by Efron about large scale inference \cite{efron2012large}.
\end{enumerate}
A limitation of our work is that have only considered the case where we have an unbiased, normally distributed estimator with known standard deviation. This is indeed restrictive, but note that the usual frequentist  inference about regression coefficients almost always relies on the (asymptotic)  normality and unbiasedness of their estimators. This is the case for linear and  generalized linear models with and without random effects, as well as for generalized estimating equations (GEE) and quantile regression \cite{stefanski2002calculus}.

The fact that the standard deviation of the estimator is typically unknown and must be estimated, can be taken into account -- to some extent -- as follows. Suppose we have observed $B=b$ and its associated two-sided $p$-value $p$. Then we can compute the absolute $z$-value as $|z|= |\Phi^{-1}(p/2)|$ and an ``implied standard error'' as $\se=|b|/|z|$. This implied value may be larger than the estimated standard error, depending on how the $p$-value was calculated.

\begin{dci}
The author declares that there is no conflict of interest.
\end{dci}

\begin{sm}

\begin{proof}
(Proposition 1) The proposition is an immediate consequence of Theorem 3.2 of Casella and Berger \cite{casella1987reconciling}.
\end{proof}

\begin{proof}
(Theorem 1) The conditional distribution of $\beta$ given $B$ is normal with mean $M=B/2$ and variance $v=\se^2/2$. Therefore,
\begin{equation*}
P(\beta > 0 \mid B)= \Phi(M/\sqrt{v}) = \Phi(B/\sqrt{2}\se).
\end{equation*}
The marginal distribution of $B$ is normal with mean $0$ and variance $2 \se^2$ and hence $B/\sqrt{2}\se$ has the standard normal distribution. By the familiar probability integral transform, it follows that $P(\beta > 0 \mid B)$ has the standard uniform distribution.
\end{proof}

\end{sm}

\bibliographystyle{SageV}

\end{document}